# Architected Various $MnO_6$ Octahedral Layers $MnO_2$ Nanostructures

R. Zeng[†, ‡], J.Q. Wang[†, ‖], G.D. Du[†, ‡], W.X. Li[†], Z.X. Chen[†], S. Li[§], Z.P. Guo[†], S.X. Dou[†]

[†]*Institute for Superconducting and Electronic Materials, School of Mechanical, Materials & Mechatronics Engineering, University of Wollongong, NSW 2522, Australia.*

[‡]*Solar Energy Technologies, School of Computing, Engineering and Mathematics, University of Western Sydney, Penrith Sout, Sydney, NSW 2751, Australia*

[‖]School of Materials Science and Engineering, University of Jinan, Jinan 250022, P. R. China.

[§]*School of Materials Science and Engineering, University of New South Wales, Sydney NSW 2502, Australia.*

Address for Correspondence:

R. Zeng

Solar Energy Technologies
School of Computing, Engineering and Mathematics
University of Western Sydney
Penrith Sout, Sydney, NSW 2751, Australia
Electronic mail: r.zeng@uws.edu.au

## Abstract

A series of $MnO_2$ unique nanostructures were synthesized by a facile hydrothermal method with microwave-assisted procedures. A novel nanostructure formation mechanism has been proposed, that $MnO_6$ octahedral nuclei and molecular layers can be rearranged into different nanostructured morphologies by restacking and splitting, and then clasping, which is associated with tunnel structure phase transformation. Systematic macrostructure observations and magnetic property measurements have been conducted on these nanostructures, which demonstrate that engineering layered structures has the potential to create many unique nanostructures and unusual physicochemical behaviours.

The engineering of layered structures has become more and more important, to the point where it has recently become a huge challenge to materials science and technology researchers, since the crystallography, electronic structure, and physicochemical properties can change significantly when a layer a few atoms thick is introduced, or even a single atomic layer thickness, such as in the case of single carbon atomic layers: graphene [1, 2]. This also applies to single molecular layers, such as $MnO_6$ octahedral layers (or manganese oxide octahedral molecular sieves) [3, 4], single $MoS_2$ molecular layers [5, 6], etc. Re-engineering thin layers in different structures could create many more unique structures and present unusual physicochemical phenomena useful for tailoring their properties for applications. Here, we present a study on engineering architectonic $MnO_6$ octahedral molecular layers by restacking thin or single $MnO_6$ layers into different tunnel structures linked with $MnO_6$ octahedra. The new molecular layers feature a series of morphologies in the forms of nanoflowers, square nanotubes, and rectangular nanowires, which are also of interest for their unusual magnetic phenomena.

Manganese dioxides with layered and tunnel structures are attractive inorganic materials owing to their distinctive physical and chemical properties, as well as their wide applications in molecular/ionic sieves,[7] catalysts, [8, 9] and electrode materials in Li/$MnO_2$ batteries. [10 - 13]. Even their unclear response mechanism has hugely improved the sensitivity of biosensors [14]. Various $MnO_2$ nanostructures with different morphologies and crystallographic forms have been reported [15-17]. $MnO_2$ has many polymorphic forms, such as $\alpha$, $\beta$, $\gamma$, $\varepsilon$, $\lambda$, and $\delta$-$MnO_2$, which are different in the way that they link the basic $MnO_6$ octahedral units [18, 19].

The formation mechanisms of the different nanostructures of $MnO_2$ have mainly been accepted as based on a rolling mechanism combined with phase transformation [15]. The phases, the morphologies, the Mn ion valences, and the microdefects all are strongly dependent on the preparation conditions (pH value, concentration of cations, and parameters). We have synthesized a series of $MnO_2$ nanostructures with different phases ($\alpha$, $\beta$, $\varepsilon$, and $\delta$) and different nano–architectonic morphologies (nanoflowers, square nanotubes, and tetragonal nanowires) by varying the preparation conditions. However, the formation mechanism has been adjusted by adding a nanoribbon stacking and restacking mechanism for our unique nanostructures under microwave hydrothermal conditions, which will be discussed below.

Magnetic nanostructures and nanomagnetism have always attracted much interest among magnetism researchers. This is due to their huge potential for technological applications in information technology [20] and in other disciplines such as biology and medicine [21]. A challenging aim of current research in magnetism is to explore structures of still lower dimensionality [22-24], and to explore the spins, orbital lattices, and couplings in the low dimension nanostructures [25-27]. As the dimensionality of a physical system is reduced, magnetic ordering tends to decrease, as fluctuations become relatively more important, but it seems that this can be overcome by engineering the surface nanostructures and step-edge atoms through introducing exchange bias [27] and enhancing the magnetic anisotropy, since step atoms present remarkably high anisotropy energy in two-dimensional nanostructures [28, 29].

In this paper, various high-quality AFM $MnO_2$ nanostructures have been synthesized, and magnetism studies with an emphasis on the relationship between the surface or interface microstructures and the magnetic properties have been performed. It was found that all the AFM $MnO_2$ nanostructures presented ferromagnetism, but with different ferromagnetic behaviors, e.g. different remnant moment and coercivity ($M_R$ and $H_C$), which is significantly enhanced by the step-edges and the surface or interface disordered clusters.

A series of $MnO_2$ nanostructures were synthesized by a facile hydrothermal method with microwave-assisted procedures based on two methods. Method A is a liquid-phase oxidation method with $Mn^{2+}$ as the main source:

$$Mn^{2+} + (NH_4)_2S_2O_8 + 2H_2O \rightarrow MnO_2 + (NH_4)_2SO_4 + 2H_2SO_4 \qquad \text{(A)}$$

Method B involves the reduction of $KMnO_4$ in a hydrochloric acid solution:

$$4KMnO_4 + 4HCl \rightarrow 4MnO_2 + 4KCl + 2H_2O + 3O_2 \ldots \qquad \text{(B).}$$

For more details see in Supporting Information (SI) in Table S1. So far, ε–nanoflowers (ε–NF), α–rectangular nanowires (α–RNW), and β–nanorods (β–NR) can be obtained by method A, while δ–nanoflowers (δ–NF), α–square nanotubes (α–SNT), and β–microcrystals (β–MC) can be obtained by method B, through tuning the microwave irradiation conditions and processing parameters.

Selected results and images of the $MnO_2$ square nanostructures are shown in Fig. 1. The x-ray diffraction (XRD) patterns of the nanostructures are shown in Fig. 1(A), where the red lines and symbols indicate the nanostructures synthesized by method A, and the blue lines and symbols indicate the ones from method B. The same lines and symbols are used in Fig. 1(B), which shows the x-ray photoelectron spectra (XPS) of the nanostructures.

The XRD patterns (Figure 1(A)) show high phase purity. All samples are single phase nanostructures under the x-ray measurement limitations. However, the XPS results shown in Fig. 1(B) and in SI of Fig. S1 indicate that the average valences of the Mn in the samples synthesized by method A are lower than those in the samples synthesized by method B, and the Mn-2p3 peaks of all the samples show splitting, which indicates that Mn exists in different valences in all the $MnO_2$ nanostructures (with the whole range of details and splitting peaks shown in the SI Fig. S1).

The field emission scanning electron microscope (FESEM) images (Fig. 2(A)-(F)) show the typical morphologies of different products: ε–NF (A), α–RNW (B), and β–NR (C), all synthesised by method A, and δ–NF (D), α–SNT (E), and β–MC (F), all synthesised by method B. They show highly uniform shapes and size distributions. Moreover, these observations of the microstructures indicate that the ε–nanoflowers have architectures composed of interconnecting nanosheets, the same as δ–nanoflowers; furthermore, the α–rectangular nanowires and the α–square nanotubes inherit their structures from a network architecture composed of rectangular nanowires and square nanotubes.

More detailed microstructures and intermediate states of nanostructure formation can be observed by transmission electron microscopy (TEM) and high resolution TEM (HRTEM), and the nanostructure formation mechanism can be obtained. Fig. 3(A) shows that the flower petals are porous nanosheets that are composed of nanoribbons and their slitted and restacked form as nanowires (as shown in inset (a) of Fig. 3(D)). Note that the δ–phase is a metastable phase as well, and the δ–nano-architecture is an intermediate state of later α–square nanotubes. Fig. 3(D) shows that the apparent flower petals (with the image of one petal shown in insets (b, c) of Fig. 3(D)), are porous nanosheets that are composed of nanoribbons and their reopened and restacked form of nanotubes. Inset (c) of Fig. 3(D) shows one such square nanotube growing out of the bottom of the image.

TEM of a single α-RNW nanowire (inset (b) of Fig. 3(B)) shows a very rough surface that hints of numerous ribbon steps and edges, and the corresponding HRTEM image (Fig. 4(B)) shows fringes of the (001) face. Selected area electron diffraction (SAED) patterns are shown in inset (c) of Fig. 3(B) and inset (a) of Fig. 4(B). The HRTEM image (Fig. 4(B)) of α-RNW shows several nanoscale clusters and fringes, which are inherent to the microstructures and defects of ε–NFs, as marked in Fig. 4(A). Moreover, the width of the nanoribbons is marked in Fig. 4(B) and runs from 5.8 nm to 9.4 nm, which can not be seen in ε–NFs. Zigzag and rhombus fringes (marked in inset (c) of Fig. 4(A)), are observed in the edges of nanoribbons. Since α-$MnO_2$ has a tetragonal Hollandite-type structure, in which the $MnO_6$ octahedra are linked to form double zigzag chains along the *c*-axis by edge-sharing, this unique crystal structure may easily form above zigzag step-edges, and, in turn, it can be the origin of the strong unusual ferromagnetism in the nanostructures. The TEM image of a single nanotube (inset (a) of Fig. 4(E)) shows a very rough surface and hints at numerous ribbon steps and edges. The HRTEM images in panels (c) and (d) of Fig. 4(E)) show fringes of the (001) face, and the corresponding electron diffraction pattern is also shown in inset (b) of Fig. 3(E)). HRTEM images (panels (c) and (d) of Fig. 4(E)) shows β–$MnO_2$-clusters and other heterostructural clusters, and a schematic diagram is presented as panel (b), based on careful observation of the four walls of a nanotube. The high resolution images of Fig. 4(E) show that the average nanotube diameter is 20 nm to 100 nm, and the average length is 500 nm, from a combined estimate from HRTEM and high resolution scanning electron microscopy (HRSEM) studies (Fig. 3(E)). These observations also indicate that the unique square nanotubes are restacked from highly oriented $MnO_2$ nanoribbons.

Many groups have investigated the synthesis methods [9-16] and the formation mechanisms [9, 16-19] of $MnO_2$ nanostructures, such as a rolling mechanism plus phase transformation. Some of them are based on template-assisted or other assisted methods [13], but the formation mechanisms are different. In the present synthesis, neither templates nor surfactants are used in the reaction system; it is only microwave assisted. Based on the above observations and analysis, we suggest that the formation of our unique nanostructures can be interpreted as based on the curling and restacking of a few simple nanoribbons, with an accompanying phase transition mechanism. The mechanism may be defined as follows: in the initial stage under microwave-assisted hydrothermal conditions, ε–phase (method A) or δ–phase (method B) $MnO_6$ octahedral nuclei are combined into nanoribbons a few atomic layers thick (see slightly curled nanoribbons in Fig. 4(D)), and hence, these layer structured nanoribons tend to curl and restack to form porous nanosheets resembling petals (see a petal image in inset (c) of Fig. 3(D)), which then form themselves into nanoflowers. Both ε–phase and δ–phase are metastable phases, and the porous nanoribbons forming nanosheets, which go on to form architectonic nanoflowers, are also in a metastable state. Under elevated temperature and pressure, or with a decreasing concentration of cations in the solution, the thin ε–nanoribbons may restack into thick nanosheets, and the thick nanosheets then split into rectangular nanowires for method A, where densely stacked ε–structure nanosheets of $MnO_6$ split and expand into 2 × 2 tunnel structure α–phase (see the initial stage of the splitting image in inset (b) of Fig. 4(A), where double arrows indicate the split sheet ends and phase transformation into twin nanograins). Similarly, the δ–nanoribbons may reopen or re-curl and restack into square nanotubes for method B, accompanied by a $MnO_6$ layer δ–structure collapse into 2 × 2 tunnel α–phase. With a longer dwell time under microwave hydrothermal conditions, the size of the nanowire or nanotube will grow larger and larger to form nanorods or a microcrystal, with an accompanying 2 × 2 tunnel structured α–phase collapse into a 1 × 1 tunnel β–phase structure.

Based the above microstructural observations and analysis, we can determine the formation mechanism of $MnO_2$ nanostructures. Fig. 5 shows the formation of $MnO_2$ nanostructures and the phase transformation mechanism from a $MnO_6$ stack of layers to tunnel structures. It should be noted that there is a slight difference during the intermediate stage between method A and method B, i.e. method A forms a thicker nanoribbon architecture that consists of metastable $MnO_6$ in densely stacked ε−phase nanoflowers, which then split and restack to 2 × 2 tunnel structure α−phase rectangular nanowires, while method B forms a much thinner nanoribbon architecture that consists of metastable $MnO_6$ in a stacked layer structure in δ−phase nanoflowers, which then curl and restack to 2 × 2 tunnel structure α−phase square nanotubes.

Due to the unique $MnO_6$ octahedral nanoribbons that are restacked into complex architectonic $MnO_2$ nanostructures, their magnetic behavior presents more unusual features. In order to elucidate those unusual magnetic phenomena, we performed a series of magnetic measurements and analyses to determine the origin of the magnetism.

Magnetic measurements (magnetization verse temperature and magnetic hysteresis loops see in SI Fig. S2 (A) and (B) respectively) show unusual and complex characteristic features. All the nanostructures show a ferromagnetic-like transition at different Curie temperatures, $T_C$, in the $MnO_2$ antiferromagnetic system. The α−RNW, β−NR, and α−SNT show a more obviously ferromagnetic transition at $T_C$ = 50 K and multiple magnetic transitions. α−SNTs show more unexpected strong ferromagnetic behaviour with remanent magnetization ($M_R$) of 0.48 emu/g and coercivity ($H_C$) of 3160 Oe; the $M_R$ and $H_C$ values of the other nanostructures are listed in SI Table SI.

We have systemically analysed the $MnO_2$ nanostructures presented in this paper. Due to their different surface or interface microstructures, they show slightly different magnetic behaviours, so that only AFM/spin glass (SG) exchange coupling behaviour was observed all $MnO_2$ nanostructures presented in this paper, which indicated that these nanostructures would appear as core-shell structural nanowires, nanotubes, nanoflowers, etc. These results will be presented elsewhere. However, the microscopic origins of magnetization (M) and coercivity ($H_C$) are similar. It is well known that disorder from element vacancies, valence changes, defects and strains, zigzag edges, and even thermal effects [20-30], etc. can result in the formation of random clusters that induce weak magnetism in nanostructures, but this can also arise from well aligned structures, such as step-edges, which can create strong magnetic anisotropy [28, 29]. All of these may exist in our $MnO_2$ nanostructures, especially at the surfaces and interfaces of the nanoribbons. Restacking of those nanoribbons leads to coexistence and competition between different magnetic behaviours and would strongly enhance their interaction exchange couplings, which generate the unusual magnetic phenomena. Since our nanostructures are all composed of different tunnel / dense packed $MnO_6$ nanosheets in different shapes, the magnetism of $MnO_2$ nanostructures seems to only originate from the surfaces or interfaces, as well as the interactions between the nanostructures, so our neutron diffraction measurements on α−SNTs do not find interesting magnetic behaviours, such as is observed by the PPMS measurements, but they thus confirm that the origin of the magnetism is from the surfaces and interfaces.

In summary, we present studies on a facile microwave-assisted hydrothermal synthesis method to prepare nanoribbons a few atoms thick, which stack into unique $MnO_2$ nanostructures, with an emphasis on the microstructures of nanoribbon surface clusters and step-edges after the restacking of the nanoribbons and on their unusual magnetic phenomena.

The formation mechanism of those $MnO_2$ nanostructures has been investigated, and a novel formation mechanism introduced, so that nanoribbons grown from nuclei stack and restack with accompanying phase transformation. Nanoribbons stacked into layers and restacked nanostructures (similar to multilayer heterostructures) present unique microstructures, and in particular, unique surface or interface microstructures and unique morphologies, and hence present unusual magnetic phenomena. In surfaces or in interfaces, the $MnO_2$ heteroclusters and step-edges, which are associated with variation of the valence of Mn ions, are the microscopic origin of the ferromagnetism, and their interactions, couplings, and competition cause the unusual ferromagnetic phenomena.

**Acknowledgments** The authors thank Dr. T. Silver for her help and useful discussions. This work is supported by the Australian Research Council.

**References**

(1) Barth, J. V.; Costantini, G. ; and Kern, K. ; *Nature* **2005**, 437, 671.
(2) Novoselov, K. S.; Geim, A. K. ; Morozov, S; Jiang, V. D.; Katsnelson, M. I. ; Grigorieva, I. V. ; Dubonos, S. V.; and Firsov, A. A.; *Nature* **2005**, 480, 10.
(3) Shen, X. F.; Zerger, R. P.; DeGuzman, R. N.; Suib, S. L.; McCurdy, L.; Potter, D. I.; and Yang, C. L. O'.; *Science,* **1993,** 260, 511.
(4) Tenne, R.; *Nat. Nanotechnol.* **2006,** 1**,** 103.
(5) Bollinger, M. V.; Lauritse, J.; Jaconsen, K. W.; Noskov, J. K.; Helveg, S.; and Besenbacher, F.; *Phys. Rev. Lett.* **2001,** 87, 196803.
(6) Jaramillo, T. F.; Jorgensen, K. P.; Bonde J.; Nielsen, J. H.; Horch, S.; and Chorkendorff, I.; *Science* **2007,** 317**,** 100.
(7) Shen, X. F.; Ding, Y. S.; Liu, J.; Han, Z. H.; Budnick, J. I.; Hines, W. A.; and Suib, S. L.; *J. Am. Chem. Soc.* **2005,** 127, 6166.
(8) Brock, S. L.; Duan, N. G.; Tian, Z. R.; Giraldo, O.; Zhou, H.; and Suib, S. L.; *Chem. Mater.* **1998**, 10, 2619.
(9) Ding, Y. S.; Shen, X. F.; Sithambaram, S.; Gomez, S., Kumar, R.; Crisostomo, V. M. B.; Suib, S. L.; and Aindow, M.; *Chem. Mater.* **2005**, 17, 5382.
(10) Armstrong, A. R.; and Bruce, P. G.; *Nature* **1996,** 381, 499.
(11) Ammundsen, B.; and Paulsen, J.; *Adv. Mater.* **2001,** 13, 943.
(12) Cheng, F.Y.; Chen, J.; Gou, X. L.; Shen, P. W.; *Adv. Mater.* **2005,** 17, 2753.
(13) Jiao, F.; Harrison, A.; Hill, A. H.; Bruce, P. G.; *Adv. Mater.* **2007,** 19, 657.
(14) Luo, X. L.; Morrin, A.; Killard, A. J.; and Smyth, M.R., *Electroanalysis* **2006,** 18, 319 – 326.
(15) Wang, X.; and Li, Y. D.; *J. Am. Chem. Soc.* **2002,** 124, 2880.
(16) Zheng, D.S.; Sun, S. X.; Fan, W. L.; Yu, H. Y.; Fan, C. H.; Cao, G. X.; Yin, Z. L.; and Song, X. Y., *J. Phys. Chem. B* **2005,** 109, 16439.
(17) Ma, R. H.; Bando, Y.; Zhang, L. Q.; and Sasaki, T.; *Adv. Mater.* **2004,** 16, 918.
(18) Thackeray, M. M.; *Prog. Solid State Chem.* **1997,** 25, 1.
(19) Sato, H.; Enoki, T.; Isobe, M.; and Ueda, Y.;. *Phys. Rev. B* **2000,** 61, 3563.
(20) Parkin, S. S. P.; Hayashi, M.; and Thomas, L.; *Science* **2008,** 320, 190.
(21) Gupta, A. K.; and Gupta, M.; *Biomaterials* **2005,** 26, 3995 ().
(22) Stamm, C., Marty, F., Vaterlaus, A., Weich, V., Egger, S., Maier, U., Ramsperger, U., Fuhrmann, H., and Pescia, D.; *Science* **1998,** 282, 449-451.
(23) Dorantes-Da Âvila, J.; and Pastor, G. M.; *Phys. Rev. Lett.* **1998,** 81, 208-211.
(24) Pratzer, M.; Elmers, H. J.; Bode, M.; Pietzsch, O.; Kubetzka, A.; Wiesendanger, R.; *Phys. Rev. Lett.* **2001,** 87, 127201.
(25) Bode, M.; Vedmedenko, E. Y.; Von Bergmann, K.; Kubetzka, A.; Ferriani, P.; Heinze, S.; and Wiesendanger, R.; *Nature Materials* **2006,** 5, 477.
(26) Prodi A., Gilioli, E., Gauzzi, A.; Licci, F.; Marezio, M.; Bolzoni, F.; Huang, Q.; Santoro, A.; Lynn, J. W.; *Nature Materials* **2004,** 3, 48.
(27) Kunes, J. et al.; *Nature Materials* 7, 198 (2008).
(28) Barth, J. V.; Costantini, G.; and Kern, K.; *Nature* **2005,** 437, 671.
(29) Gambardella, P.; Rusponi, S.; Veronese.; M, Dhesi, S. S.; Grazioli, C.; Dallmeyer, A.; Cabria, I.; Zeller, R.; Dederichs, P. H.; Kern, K.; Carbone, C.; Brune, H.; *Science* **2003,** 300, 1130.
(30) Benitez, M. J., Petracic, O.; Salabas, E. L.; Radu, F.; Tuysuz, H.; Schuth, F.; Zabel, H.; *Phys. Rev. Lett.* **2008,** 101, 097206.

# Figure Captions

**Figure 1.** **(A)** X-ray diffraction patterns and **(B)** selected XPS spectra of $MnO_2$ nanostructures: ε-$MnO_2$ nanoflowers (ε-NF), α-$MnO_2$ rectangular nanowires (α-RNW), β-$MnO_2$ nanorods (β-NR), δ-$MnO_2$ nanoflowers (δ-NF), α-$MnO_2$ square nanotubes (α-SNT), and β-$MnO_2$ microcrystals (β-MC).

**Figure 2.** FESEM images of $MnO_2$ nanostructures: **(A)** ε-NF, **(B)** α-RNW, **(C)** β-NR, **(D)** δ-NF, **(E)** α-SNT, and **(F)** β-MC.

**Figure 3.** TEM images of $MnO_2$ nanostructures: **(A)** ε-NF, with inset of the corresponding SAED pattern; **(B)** α-RNW, with inset (a) a magnified image, inset (b) the SAED pattern, and inset (c) an HRTEM image of a single nanotube; **(C)** β-NR, with inset of the corresponding SAED pattern; **(D)** δ-NF, with the insets showing magnified details at an intermediate state of tube formation: TEM image (a), single nanotube images in TEM (b), and HRTEM (c); and **(E)** HRTEM image of the α-$MnO_2$ square nanotubes, with inset (a) showing the HRTEM image of a single nanotube and inset (b) the corresponding SAED pattern.

**Figure 4.** HRTEM images of $MnO_2$ nanostructures: **(A)** ε-NF, **(B)** α-RNW, **(C)** β-NR, **(D)** δ-NF, and **(E)** α-SNT. The average diameter of the wires is 20-50 nm, and the average length is 500 nm. Details of the insets and composite panels are given in the text.

**Figure 5.** Schematic of the $MnO_2$ nanostructure formation process and phase transformation mechanism from a stack of $MnO_6$ layers to tunnel structures produced by restacking under microwave-assisted hydrothermal conditions.

# Figures

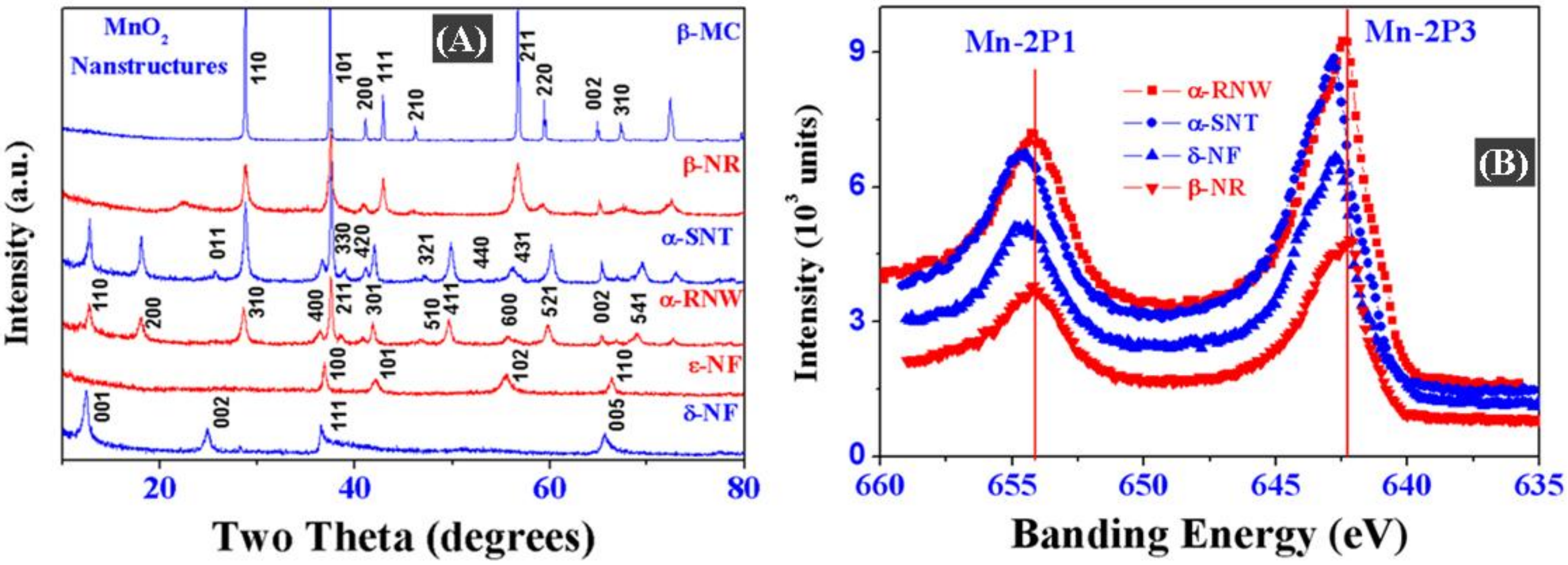

Figure 1, R. Zeng et al.

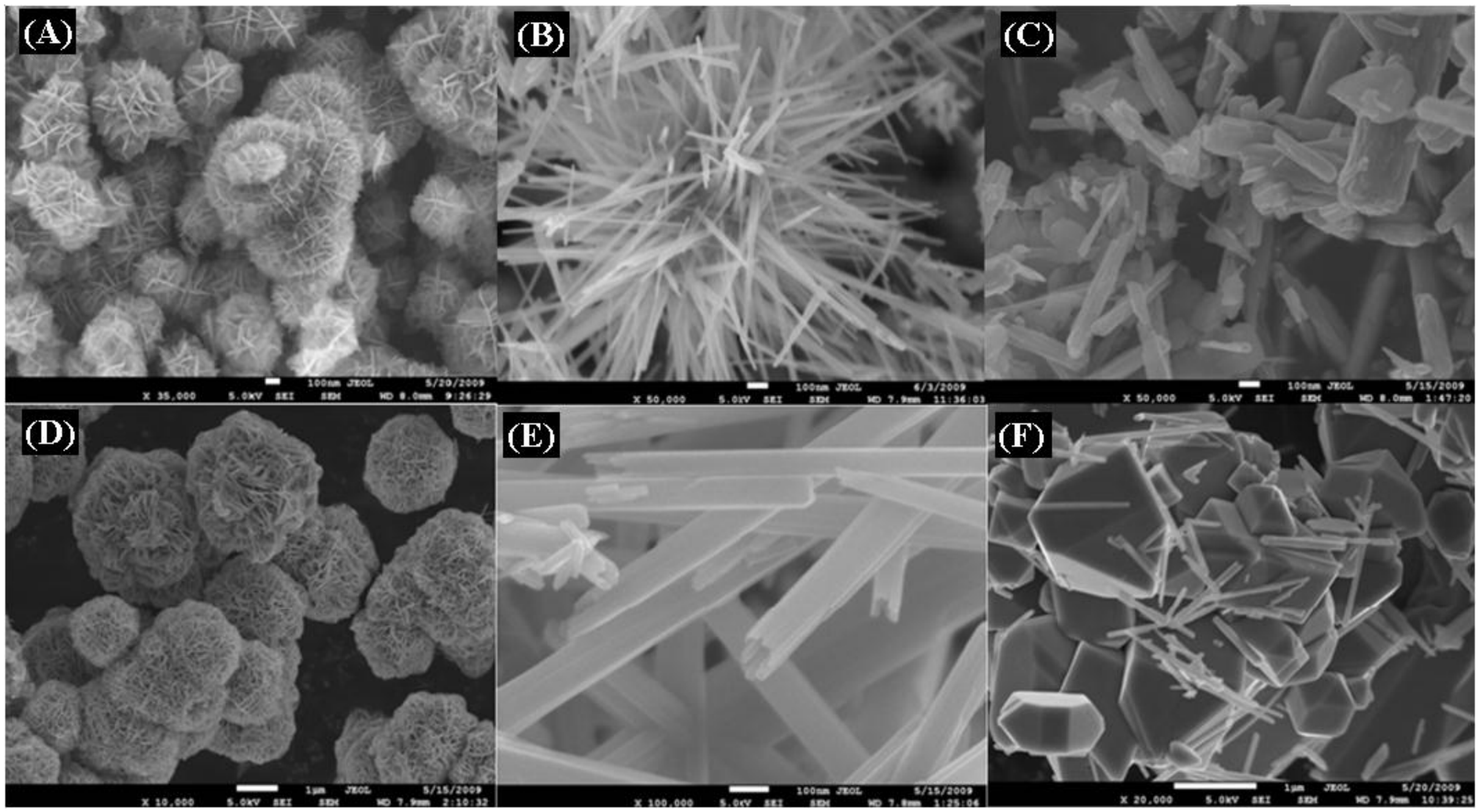

Figure 2, R. Zeng et al.

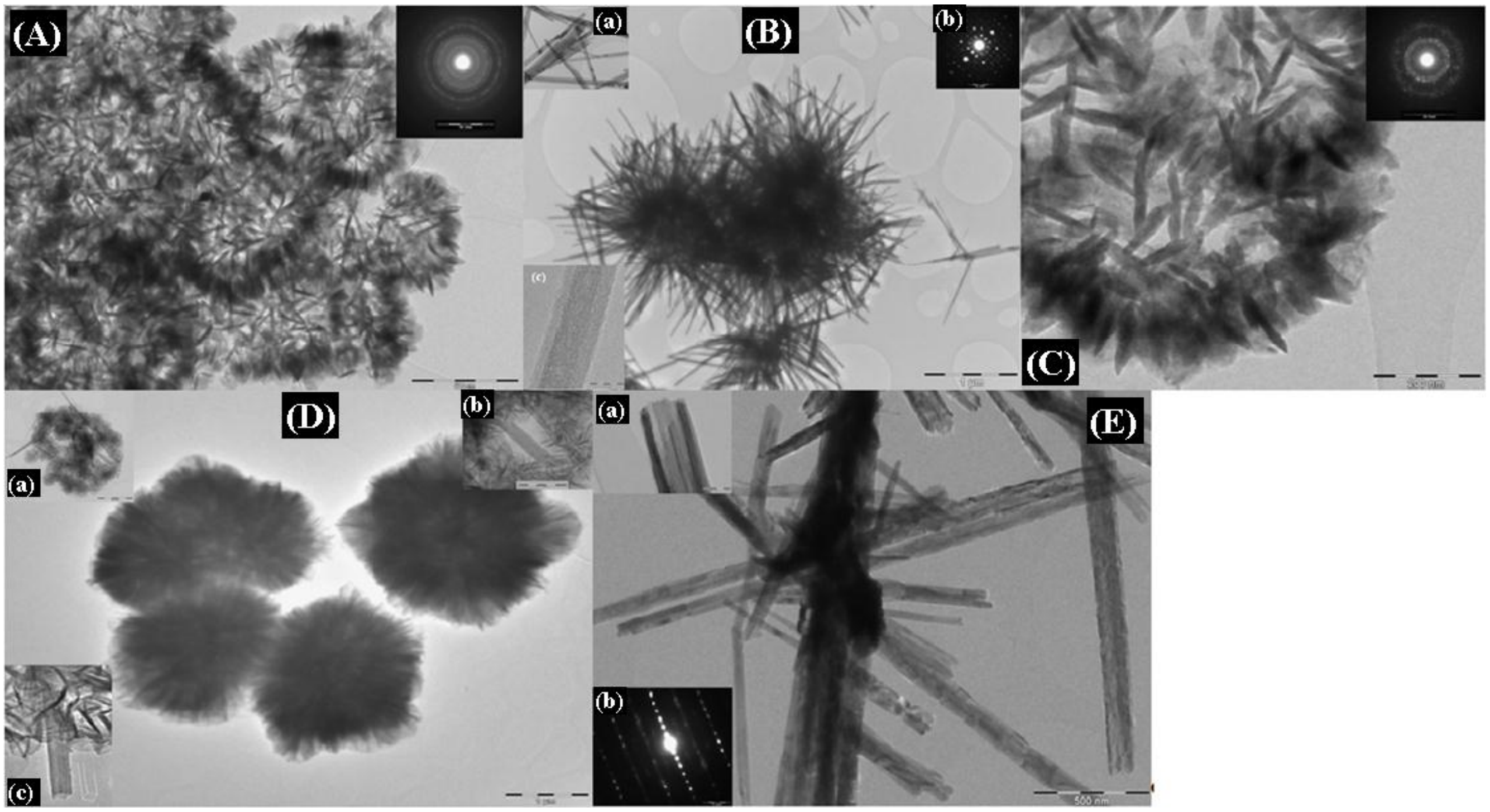

Figure 3, R. Zeng et al.

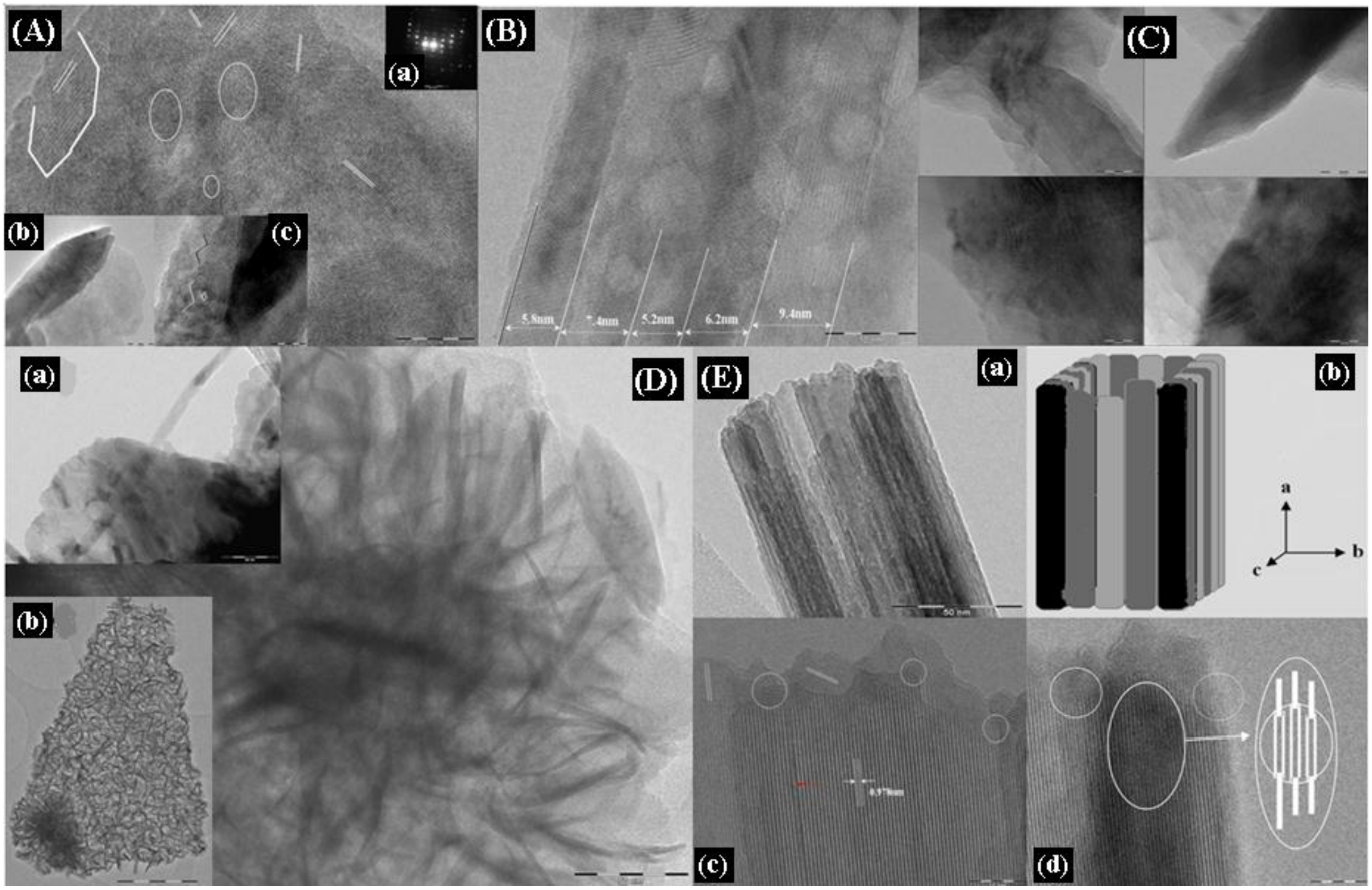

Figure 4, R. Zeng et al.

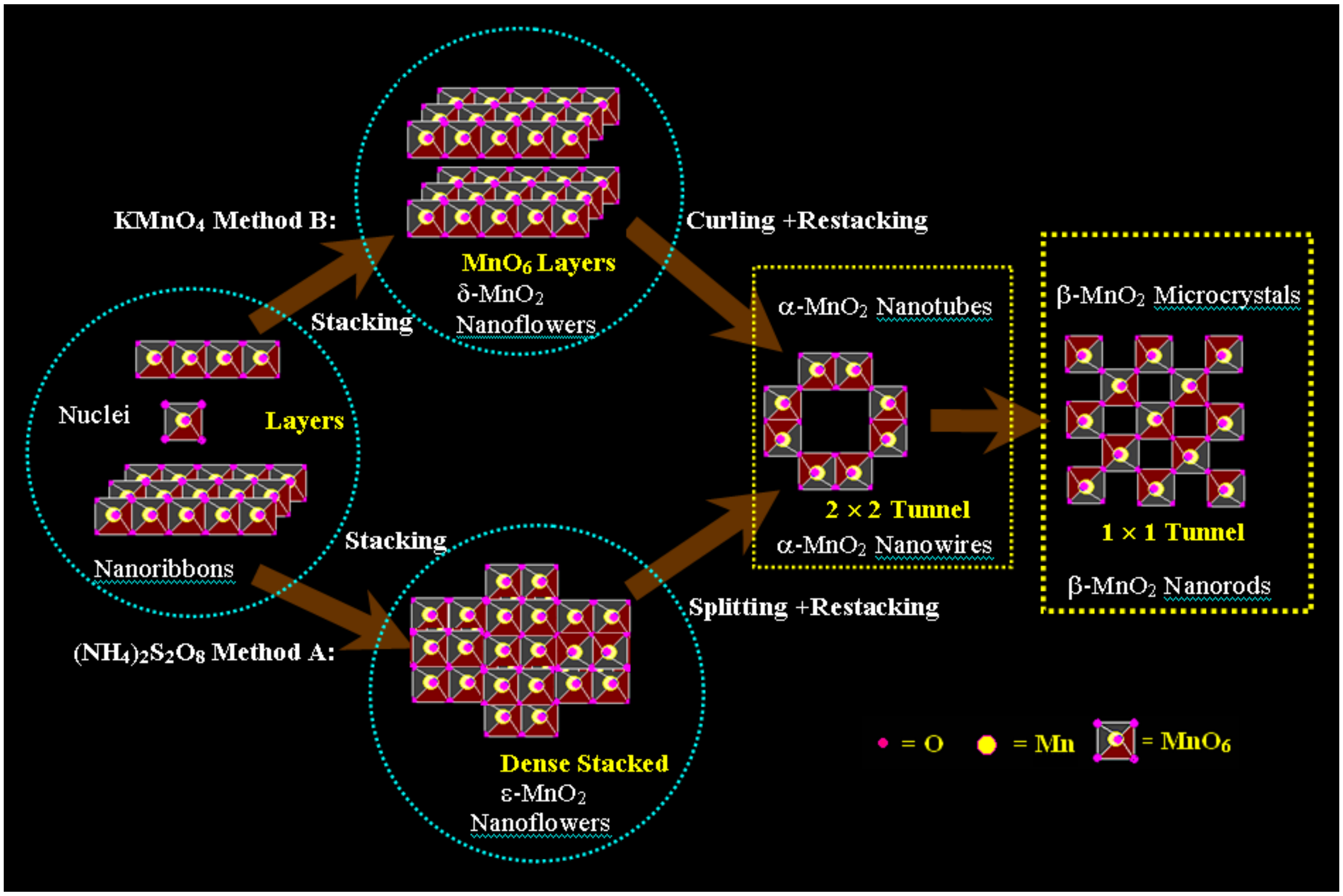

Figure 5, R. Zeng et al.

# Supporting Information for

# Various $MnO_6$ Octahedral Layers Architectonic $MnO_2$ Nanostructures

R. Zeng[†, ‡], J.Q. Wang[†, ‖], G.D. Du[†, ‡], W.X. Li[†], Z.X. Chen[†], S. Li[§], Z.P. Guo[†], S.X. Dou[†]

[†]*Institute for Superconducting and Electronic Materials, School of Mechanical, Materials & Mechatronics Engineering, University of Wollongong, NSW 2522, Australia.*

[‡]*Solar Energy Technologies, School of Computing, Engineering and Mathematics, University of Western Sydney, Penrith Sout, Sydney, NSW 2751, Australia*

[‖]School of Materials Science and Engineering, University of Jinan, Jinan 250022, P. R. China.

[§]*School of Materials Science and Engineering, University of New South Wales, Sydney NSW 2502, Australia.*

Address for Correspondence:

R. Zeng

Solar Energy Technologies
School of Computing, Engineering and Mathematics
University of Western Sydney
Penrith Sout, Sydney, NSW 2751, Australia
Electronic mail: r.zeng@uws.edu.au

In particular, nanostructures consisting of an antiferromagnetic (AFM) material have been of the greatest interest in recent years [1-8]. As the size of a magnetic system decreases, the importance of the surface and its roughness or surface step atom quantum effects increases. Since an antiferromagnet usually has two mutually compensating sublattices, the surface always leads to a breaking of the sub-lattice pairing and thus to ‘‘uncompensated’’ surface spins. This effect has already been explained as the origin of exchange-bias and net magnetic moment in AFM nanoparticles by Néel [2].

It is well known that most forms of $MnO_2$ present AFM behaviours, and α-$MnO_2$ is anti-ferromagnetic, so a reduction in the oxidation state of the Mn mixed valence of $Mn^{3+}$ and $Mn^{4+}$ is necessary in order to compensate for the charge of introduced large cations [1], which could influence the magnetic coupling between the Mn cations. On the other hand, the distribution of the $Mn^{3+}$ and $Mn^{4+}$ cations should be closely related to the distribution of the intercalated cations, which in turn, may cause a change in the magnetic ground state. In addition, the tetragonal rutile-type β-$MnO_2$ is the thermodynamically most stable and abundant member of the manganese dioxide family, and it plays an important role in magnetism and transport properties [3, 4]. β-$MnO_2$ shows a magnetic transition into a helical state at the Néel temperature ($T_N$) of about 92 K, below which it has a well-known screw type incommensurate magnetic structure, with the pitch of the screw about 4% shorter than 7/2*c* [4]. Above $T_N$, the magnetoresistance (MR) of β-$MnO_2$ is slightly negative and isotropic. However, below $T_N$, on the other hand, the MR becomes anisotropic and remains small. In this report, we focus on the magnetic properties of magnesium dioxide. In this paper, various high-quality AFM $MnO_2$ nanostructures have been synthesized, and magnetism studies with an emphasis on the relationship between the surface or interface microstructures and the magnetic properties have been performed. It was found that all the AFM $MnO_2$ nanostructures presented ferromagnetism, but with different ferromagnetic behaviors, e.g. different remnant moment and coercivity ($M_R$ and $H_C$), which is significantly enhanced by the step-edges and the surface or interface disordered clusters.

Table S1. Summary of the synthesis conditions, crystallographic structure, morphology, and properties of $MnO_2$ nanostructures synthesized by a microwave-assisted hydrothermal method.

| Samples | Synthesis conditions | Crystallographic Structures | Morphology | Magnetic Properties (at 5K, $H_{ZFC}$=7T) | | |
|---|---|---|---|---|---|---|
| | | | | $M_R$ (emu/g) | $M_s$ (emu/g) | $H_C$ (Oe) |
| ε – NF | $Mn^{2+}$ sources + $(NH_4)_2S_2O_8$ solution reaction method. 110°C for 1h | ε-$MnO_2$: hexagonal phase, space group:P63/mmc(194), lattice constants *a*=*b*=2.85Å, *c*=4.65 Å, dense stack | Nanoribbons composed of dense packed $MnO_6$ nanosheets restacked into architectural porous nanoflowers | 0.23 | 15.4 | 225 |
| α – RNW | $Mn^{2+}$ sources + $(NH_4)_2S_2O_8$ solution reaction method. $H_2SO_4$ 140°C for 1h | α-$MnO_2$, tetragonal phase, space group:I4/m(87), lattice constants *a*=*b*=9.865Å, *c*=2.897 Å, 2 × 2 tunnel | Nanoribbons composed of 2 × 2 tunnel packed $MnO_6$ restacked into rectangular nanowires | 0.1 | 16.8 | 654 |
| β – NR | $Mn^{2+}$ sources + $(NH_4)_2S_2O_8$ solution reaction method. 140°C for 3h | β-$MnO_2$, tetragonal phase, space group:P42/mnm(136), lattice constants *a*=*b*=4.45Å, *c*=2.93 Å, 1 × 1 tunnel | Nanoribbons composed of 1 × 1 tunnel packed $MnO_6$ restacked into rectangular nanowires | 0.06 | 5.6 | 1038 |
| δ – NF | The reduction of $KMnO_4$ method. 110°C for 1h | δ-$MnO_2$ rhombohedral phase, space group:R-3m(166), lattice constants *a*=*b*=2.94Å, *c*=21.86 Å, $MnO_6$ layer structure | Nanoribbons composed of packed $MnO_6$ nanosheets restacked into architectural porous nanoflowers | 0.05 | 15.4 | 147 |
| α – SNT | The reduction of $KMnO_4$ method. 140°C for 1h | α-$MnO_2$ tetragonal phase, space group:I4/m(87), lattice constants *a*=*b*=9.85Å, *c*=2.86 Å, 2 × 2 tunnel | Nanoribbons composed of 2 × 2 tunnel packed $MnO_6$ restacked into square nanotubes | 0.48 | 8.6 | 3160 |
| β – MC | The reduction of $KMnO_4$ method. 140°C for 3h | β-$MnO_2$ tetragonal phase, space group:P42/mnm(136), lattice constants *a*=*b*=4.39Å, *c*=2.86 Å. 1 × 1 tunnel | Microsized crystals | N/A | N/A | N/A |

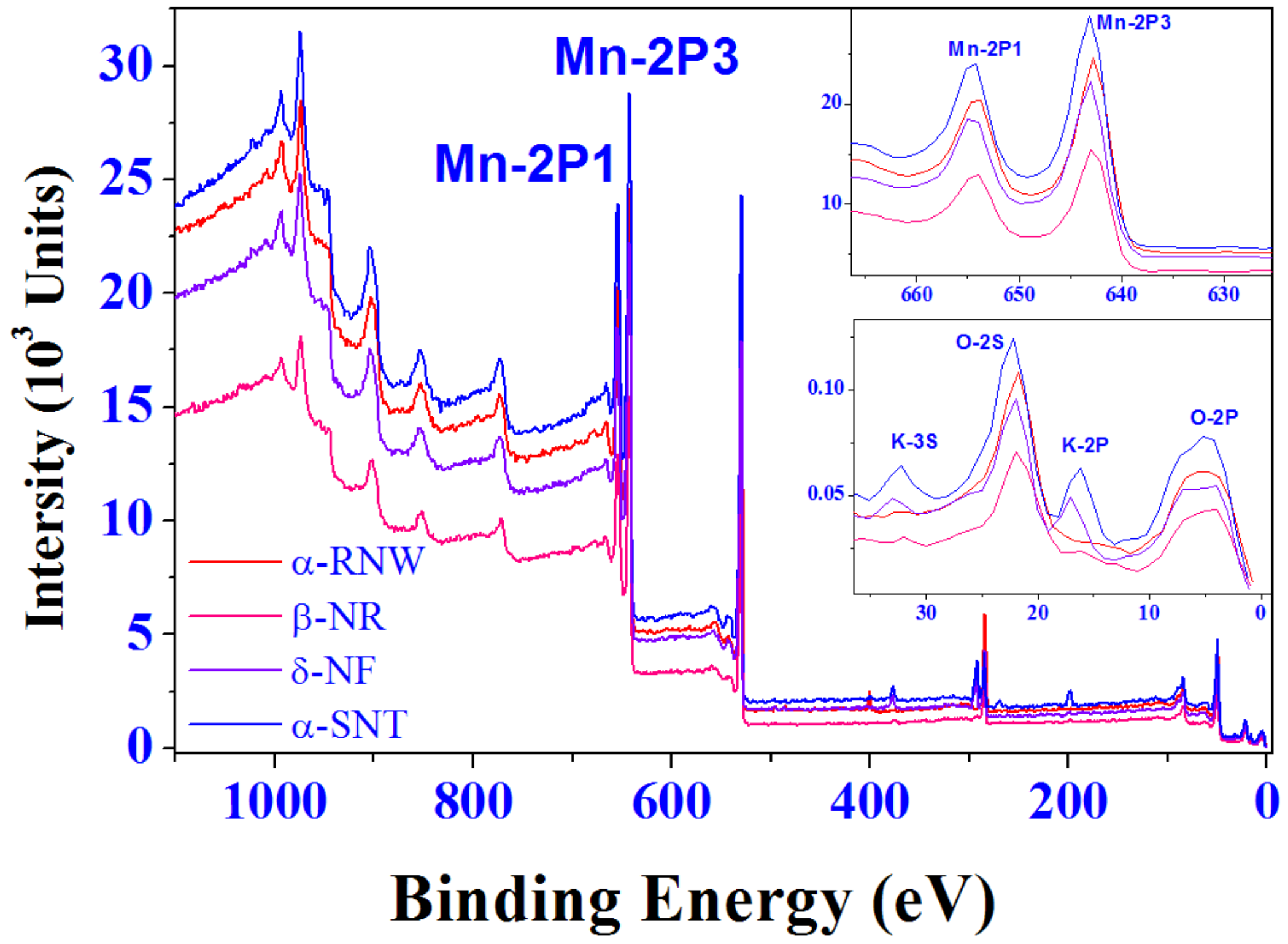

**Figure S1.** Selected XPS spectra of $MnO_2$ nanostructures: α-$MnO_2$ rectangular nanowires (red line), β-$MnO_2$ nanorods (pink line), δ-$MnO_2$ nanoflowers (violet line), and α-$MnO_2$ square nanotubes (blue line).

Inset (a) is an enlargement of the Mn-2P1 and Mn-2P3 peaks, and inset (b) is an enlargement of the O-2S, O-2P and K-3S, K-2P peaks.

The M-T curves after field cooling (FC), measured in an applied field of 50 Oe, and the M - H hysteresis loops of the $MnO_2$ nanostructures, measured at 5 K after zero-field cooling (ZFC) and with an applied field up to 70 kOe, are shown in Fig. S2 (A) and (B) respectively.

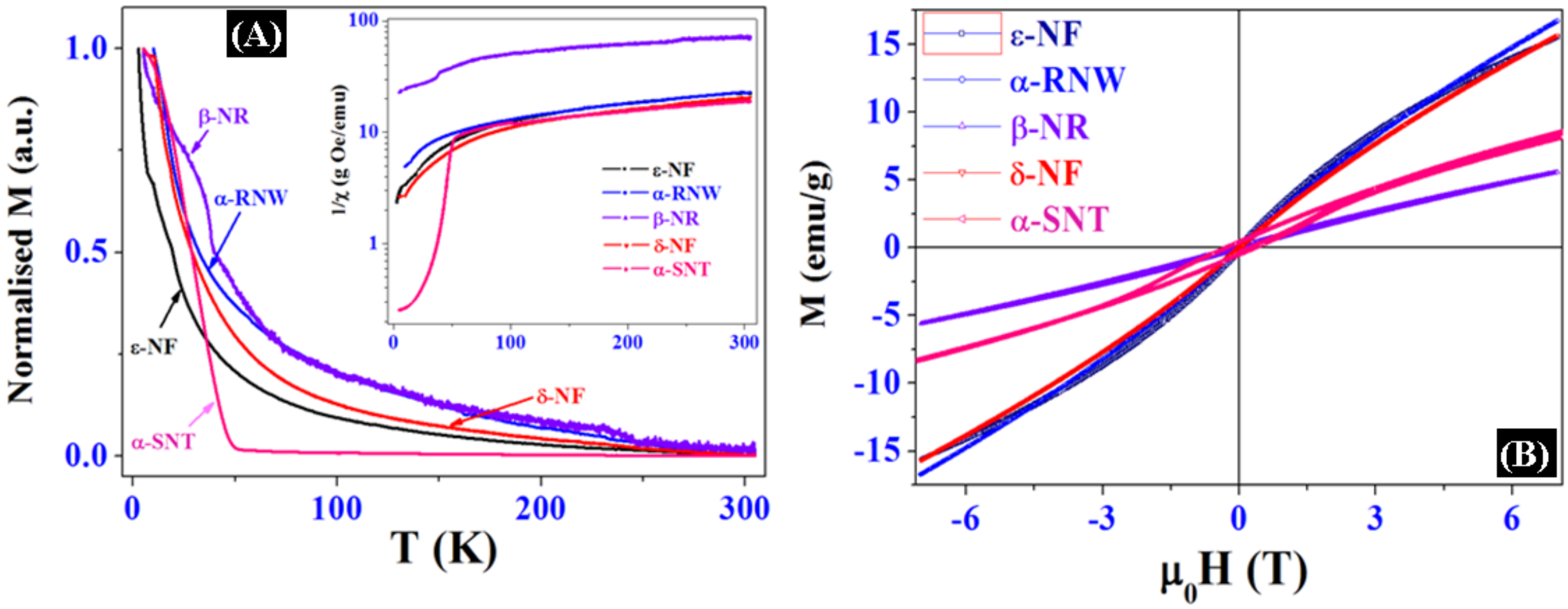

**Figure S2. (A)** M vs. T curves after field cooling (FC), measured under a field 50 of kOe (1); the inset contains 1/χ (= H/M) – T curves after FC under a field of 50 kOe. **(B)** M vs. H curves at 5 K after zero field cooling (ZFC) of $MnO_2$ nanostructures.

The induced magnetism of the clusters should first present superparamagnetic (SPM) behavior, and SPM really does have a role in the above demonstrated magnetic behavior in α–SNTs. However, the correlation between the SPM behavior of $MnO_2$ that is present in the nanostructures and their sizes is not consistent with the size dependence of magnetic anisotropy in magnetic nanoparticles according to the Stoner-Wohlfarth theory [8, 10]. This is because the microscopic origin of the magnetic anisotropy of the nanostructures stacked from nanoribbons is different from that of the corresponding nanoparticles and is mainly dependent on the step-edges. The magnetic anisotropy is an energy barrier to prevent magnetization from varying from one direction to the other. The blocking temperature is the threshold point of thermal activation to overcome such a magnetic anisotropy and to transfer magnetic nanostructures to the superparamagnetic state. A larger amount of step-edges implies higher magnetic anisotropy energy, and consequently, a higher thermal energy is required for nanostructures to become superparamagnetic. The coercivity ($H_C$) represents the required strength of the magnetic field to overcome the magnetic anisotropy barrier and to allow the magnetization of nanocrystals to align along the field direction. The coercivity of a magnetic nanocrystal from the Stoner- Wohlfarth theory can be expressed as:

$$H_C = 2K / \mu_0 M_S,$$

where $\mu_0$ is the universal constant of permeability in free space, K is the crystal anisotropy constant and $M_S$ is the saturation magnetization of the nanoparticles. When the temperature is below the blocking temperature for the given nanocrystals, the required coercivity for switching the magnetization direction of the nanocrystals certainly increases as the magnetic anisotropy increases. Therefore, the coercivity of the α-$MnO_2$ nanostructures increases with increasing amounts of step-edges. Comparing the magnetization loops of different $MnO_2$ nanostructures in Fig. 7(B) and the $M_S$, $M_R$, and $H_C$ values listed in Table I, the $H_C$ reaches the highest value in α−SNT, which is in agreement with the HRTEM observations, as there are very coarse surfaces and a large amount of step-edges existing on both the outside and the inside surfaces of the nanotubes.

**References**

(1) Shen, X. F.; Ding, Y. S.; Liu, J.; Han, Z. H.; Budnick, J. I.; Hines, W. A.; and Suib, S. L.; *J. Am. Chem. Soc.* **2005,** 127, 6166.

(2) Benitez, M. J., Petracic, O.; Salabas, E. L.; Radu, F.; Tuysuz, H.; Schuth, F.; Zabel, H.; *Phys. Rev. Lett.* **2008,** 101, 097206.

(3) Neel, L.; *Comptes Rendus,* **1961,** 252, 4075.

(4) Thackeray, M. M.; *Prog. Solid State Chem.* **1997,** 25, 1;

(5) Kovalev, O.V.; *Low Temp. Phys.* **1999,** 25, 115.

(6) Luo, J.; Zhu, H. T.; Zhang, F.; Liang, J. K.; Rao, G. H.; Li, J. B.; Du, Z. M.; *J. Appl. Phys.* **2009,** 97, 105093925.

(7) Regulski, M.; Przenioslo, R.; Sosnowska, I.; Hoffmann, J.-U.; *Phys. Rev. B* **2003,** 68, 172401.

(8) Golosovsky, I.V.; Salazar-Alvarez, G.; Lopez-Ortega, A.; Gonzalez, M. A.; Sort, J.; Estrader, M.; Surinach, S.; Baro, M. D; and Nogues J.; *Phys. Rev. Lett.* **2009,** 102, 247201.

(9) Malozemoff, A. P.; *Phys. Rev. B* **1987,** 35, 3679; *J. Appl. Phys.* **1988,** 63, 3874.

(10) Stoner, E. C.; and Wohlfarth, E. P.; *Trans. R. Soc. London, Ser. A* **1948,** 240, 599; and Stoner, E. C.; and Wohlfarth, E. P. A.; *IEEE Trans. Magn.* **1991,** 27, 3475.